\documentclass[prb,aps,twocolumn,amsmath,amssymb,floatfix,
superscriptaddress]{revtex4}
\usepackage[dvips]{graphics}
\usepackage{color}
\definecolor{dred}{rgb}{0.75,0,0}
\usepackage{soul}
\usepackage[colorlinks=true, citecolor=blue, urlcolor=blue ]{hyperref}
\date{\today}

\begin{document}

\title{Magnetic response of interacting electrons in a spatially 
non-uniform disordered multi-channel system: Exact and mean-field results}

\author{Arpita Koley}

\affiliation{Physics and Applied Mathematics Unit, Indian Statistical
Institute, 203 Barrackpore Trunk Road, Kolkata-700 108, India}

\author{Santanu K. Maiti}

\email{santanu.maiti@isical.ac.in}

\affiliation{Physics and Applied Mathematics Unit, Indian Statistical
Institute, 203 Barrackpore Trunk Road, Kolkata-700 108, India}

\begin{abstract}

In this work we explore magnetic response of interacting electrons in a spatially non-uniform disordered system, where impurities
are introduced in one sector of the geometry keeping the other one free. The interaction among the electrons are taken in the well known
Hubbard form which leads to several anomalous features in energy spectra and flux driven circular current, depending on the concentration
of up and down spin electrons. For smaller systems with less number of electrons we present exact results which always give a clear picture
to understand the basic mechanisms, while for large systems having higher number of electrons mean-field results are given. The effect 
of disorder is very interesting. Beyond a critical disorder, completely contrasting signature is obtained compared to uniform disordered 
systems, and the phenomenon becomes more promising when the Hubbard interaction is included. Along with these features, we find unusual 
half flux-quantum periodic current at some typical cases and a suitable explanation of it is provided. Our detailed analysis may be 
utilized to study magnetic response of interacting electrons in other similar kind of non-uniform disordered systems.

\end{abstract}

\maketitle

\section{Introduction}

Persistent current is a well known phenomenon in the field of mesoscopic physics. When an isolated conducting loop is threaded by a
magnetic flux, commonly referred as Aharonov-Bohm (AB)~\cite{ab1,ab2} flux, a net current appears. Once the current is established, it 
never dissipates to zero even in presence of impurities. This fact was first successfully proposed by B\"{u}ttiker, Imry and Landauer in 
their seminal work~\cite{pc1} during the early 80's and soon after this proposition interest in this area has rapidly grown up
with lots of theoretical and experimental works~\cite{pc2,pc3,pc4,pc5,pc6,pc7,pc8,pc9,pc10,pc11,pc12,pc13,pc14,pc15,pc16,pc17,liang,chen}.

Till date an enormous amount of work has been done exploring several interesting features, though some open questions still persist 
even today. For example, {\em the interplay between electron-electron (e-e) interaction and random disorder in a 
`spatially non-uniform' disordered material has not been addressed}, that might be an interesting one. Traditionally impurities 
are included uniformly~\cite{pc18,pc19,pc20,pc21} throughout the sample and under this situation the role of impurities are well 
studied both for the non-interacting~\cite{pc2,pc11} and interacting electron~\cite{pc22,pc23,pc24} cases. But no attempt has been 
made considering an ordered-disordered separated (ODS) system where one part is completely free and the other part is subjected 
to impurities~\cite{pc25,pc26,pc27}. The recent breakthroughs in nanoscience and nanotechnology allow us to get tailor made ODS systems 
in different geometrical shapes like nanotubes, thick rings, 2D surfaces and to name a few. Apparently it seems that the role of 
disorder in such systems will be quite analogous to what we observe in the case of fully disordered ones i.e., the Anderson type 
localization~\cite{pc18,pc19,anlc1,anlc2,anlc3,anlc4}. Though the situation is not like that, as pointed out by Zhong and 
Stocks~\cite{pc25} in their work. 
Considering a shell-doped nanowire they have shown that beyond a critical disorder, electrons get delocalized and it becomes more 
effective as we increase the disorder strength. A sharp localization-to-delocalization transition takes place in such an ODS system, 
and importantly, the mobility edge (ME)~\cite{skmprl} that appears is highly robust and it does not vanish even in the limit of strong 
disorder. This is in complete contrast to the 3D random disordered lattices where the ME vanishes at large disorder. Thus, in the context 
\begin{figure}[ht]
{\centering \resizebox*{4cm}{5cm}{\includegraphics{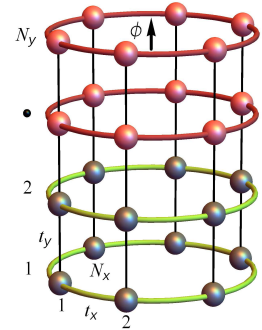}}\par}
\caption{(Color online). ODS nanotube threaded by an AB flux $\phi$, where the filled colored balls represent the atomic sites. Impurities 
are introduced in the sites shown by the light red balls, while the sites denoted by the dark gray balls are free from impurities. 
A net circular current appears in the tube due to the flux $\phi$.}
\label{fig1}
\end{figure}
of electronic localization, an ODS system is a very good example that may provide some non-trivial features, especially when we include
the effect of electron-electron correlation.

The other open question which is associated with the phenomenon of persistent current, to the best of our concern, is the lack of 
suitable explanation of getting half flux-quantum periodicity $\phi_0/2$ ($\phi_0=h/e$, the elementary flux-quantum where $e$ and $h$ 
are the fundamental constants) with AB flux $\phi$. Usually, the current provides $\phi_0$ periodicity and this issue is well 
understood. But, for some typical cases depending on the filling factor, the current exhibits $\phi_0/2$ periodicity, and its specific 
reason is not fully clear. Thus, further probing is definitely required for the sake of completeness. Though few attempts have been
made~\cite{pc6,pc28,new1,new2}, here we intend to explain the phenomenon of half flux-quantum periodicity in other way. For the ordered 
nanotube, we substantiate it analytically. An analytical treatment is always helpful to understand any phenomenon quite easily. For the 
disordered cases, numerical results are given.

The present work mainly focuses on two issues. (i) The interplay between e-e interaction and impurities in an ODS nanotube 
(see Fig.~\ref{fig1}), where the e-e correlation is included in the form of well known Hubbard model~\cite{pc29,pc30,pc31}. (ii) The 
possible reason of getting half flux-quantum periodic persistent current. Now, working with a many-body system has always been a 
challenging task as the main difficulty arises in finding the electronic energy eigenvalues. In our work, we proceed in two ways. 
For small size systems, we compute the energy eigenvalues by diagonalizing the full many-body Hamiltonian. Whereas, this prescription 
is no longer suitable for bigger systems, and thus, Hartree-Fock (HF) mean-field (MF) prescription~\cite{pc30,pc31,pc32,pc33,pc34} 
is employed. The interplay between disorder and e-e interaction is really interesting which we examine critically. Along with this, 
we give a suitable mathematical description of getting half flux-quantum periodicity in persistent current for the ordered case. 

Following the above introduction, in Sec. II, we describe briefly the nanotube and its Hamiltonian. Section III contains the theoretical 
prescription for calculating the results. All the numerical results are thoroughly discussed in Sec. IV. A brief outline about the 
experimental feasibility of nano rings is given in Sec. V, for the sake of completeness. Finally, we conclude in Sec. V.

\section{Nanotube and the Hamiltonian}

The sketch of the nanotube is shown in Fig.~\ref{fig1} in which $N_y$ number of rings are stacked vertically, where each of these rings 
contains $N_x$ number of lattice sites. We introduce random impurities in the upper $N_y/2$ ($N_y$ to be considered as even) rings,
keeping the other half free, to make the tube as an ordered-disordered separated one. The tube encloses an AB flux $\phi$, measured in 
units of $\phi_0$, which produces a net circular current in the system. 

We use a tight-binding (TB) framework to describe the system. Under nearest-neighbor hopping (NNH) approximation, the 
TB Hamiltonian of the system in presence of Hubbard interaction reads as~\cite{pc29,pc30}
\begin{eqnarray}
H &=&\sum_{j,k,\sigma} \epsilon_{j,k,\sigma} c_{j,k,\sigma}^{\dagger} c_{j,k,\sigma} + t_y
\sum_{j,k,\sigma} c_{j,k,\sigma}^{\dagger} c_{j+1,k,\sigma} \nonumber \\
& + & t_x \sum_{j,k,\sigma} e^{i\theta} c_{j,k,\sigma}^{\dagger} c_{j,k+1,\sigma} + h.c. \nonumber \\ 
& + & U\sum_{j,k} c_{j,k,\uparrow}^{\dagger} c_{j,k,\uparrow} c_{j,k,\downarrow}^{\dagger} c_{j,k,\downarrow}
\label{eqn1}
\end{eqnarray}
where the integers $j$ (runs from $1$ to $N_y$) and $k$ (varies from $1$ to $N_x$) are used to refer the ring index and the lattice sites 
in each of these rings, respectively. $\epsilon_{j,k,\sigma}$ denotes the on-site energy at a particular site ($j,k$) of an electron having 
spin $\sigma$ ($\uparrow,\downarrow$), and $U$ gives the on-site Hubbard interaction strength. $t_x$ and $t_y$ correspond to the intra-ring
and inter-ring NNH integrals, respectively, and $i=\sqrt{-1}$. Due to the AB flux $\phi$, a phase factor $\theta$ is 
introduced~\cite{pc22,deo,kus,kulik,sai} whenever an electron hops from one site of a ring to its neighboring site, and it is expressed 
as~\cite{pc22,deo,kus,kulik,sai} $\theta=2\pi \phi/N_x\phi_0$. For our system we assume that the flux $\phi$ threads the tube axially 
in such a way that the magnetic field becomes zero at the ring circumference. Therefore, the electrons always move in the field free 
space, and under this condition the electronic states depend only on the net flux (viz, $\phi$) penetrated by individual rings in the 
tube~\cite{pc2}. In traversing through a particular ring path even though the magnetic field is zero, the electronic wave function picks 
up a phase, and for one complete rotation the phase factor is $2\pi \phi/\phi_0$. This is the well known Aharonov-Bohm 
effect~\cite{pc2,deo}. As there are $N_x$ bonds in individual rings, the phase factor associated in each hopping becomes 
$\theta=2\pi \phi/N_x\phi_0$. No such phase factor appears during the motion along the vertical direction, since for this motion no flux 
is enclosed. For the ordered sites, the site energies are uniform and we fix them to zero, without loss of generality. 
While they are chosen `randomly'~\cite{pc18,pc19,pc22,kus} from a `Box' distribution function of width $W$ within the range $-W/2$ to 
$W/2$. $c_{j,k,\sigma}$ and $c_{j,k,\sigma}^{\dagger}$ are the usual fermionic operators. 
In our work we assume that the hopping integrals do not depend of the spin index ($\sigma$), and also the Hubbard interaction is 
uniform throughout sample. Accordingly, we drop spin label $\sigma$ from these parameters.

\section{Theoretical framework}

For the above TB Hamiltonian Eq.~\ref{eqn1} we investigate the characteristic features of interacting electrons. To explore the behavior 
of flux-driven circular current, first we need to determine the energy eigenvalues, which we compute in two ways. For the case of 
exact diagonalization of full many-body Hamiltonian matrix, within our computational capacity, we can reach the maximum limit of six 
electrons with three up and three down spins, and the total number of lattice sites that can be considered is ten. For larger systems 
with higher number of electrons we use HF MF scheme. Below we discuss in detail both the two prescriptions for the benefit of the readers.

\subsection{Exact diagonalization of the full many-body Hamiltonian}

The construction of the full many-body Hamiltonian is one of the key parts for theoretical calculations, and once it is done we find 
the energy eigenvalues by diagonalizing the Hamiltonian. Here we give an example of forming such a Hamiltonian considering a simple 
system, which is a $N_x$-site single channel ring ($N_y=1$) with two opposite spin electrons, viz, $N_{\uparrow}=N_{\downarrow}=1$ 
($N_{\uparrow}$ and $N_{\downarrow}$ are used to refer the total number of up and down spin electrons in a system). It can easily be 
generalized for multi-ring systems with higher number of electrons. For the single ring, we can skip the strand index $j$ in order to 
write the TB Hamiltonian in a simplified form and it is expressed as
\begin{eqnarray}
H_R &=&\sum_{k,\sigma} \epsilon_{k,\sigma} c_{k,\sigma}^{\dagger} c_{k,\sigma} 
+t_x \left(\sum_{k,\sigma} e^{i\theta} c_{k,\sigma}^{\dagger} c_{k+1,\sigma} + h.c.\right) \nonumber \\ 
& + & U \sum_k c_{k,\uparrow}^{\dagger} c_{k,\uparrow} c_{k,\downarrow}^{\dagger} c_{k,\downarrow} \nonumber \\
& = & H_1 + H_2 + H_3
\label{eqn2}
\end{eqnarray}
where the sub-Hamiltonians $H_1$, $H_2$ and $H_3$ are associated with the site energy, NNH and the Hubbard interaction terms, respectively. 
Our aim is to find out the matrix elements $\langle \eta|H|\zeta\rangle$, where $|\eta\rangle$ and $|\zeta\rangle$ are the many-particle
Wannier states. The general forms of these states are:
$$
|\eta\rangle = c_{p,\uparrow}^{\dagger} c_{q,\downarrow}^{\dagger}|0\rangle 
~~\mbox{and} ~~
|\zeta\rangle = c_{r,\uparrow}^{\dagger} c_{s,\downarrow}^{\dagger}|0\rangle
$$
where $|0\rangle$ represents the null state. $p$, $q$, $r$ and $s$ are the integers and they run from $1$ to $N$. Here it is important to 
note that, for higher number of up and/or down spin electrons we need to consider the values of these integers selectively satisfying the 
Pauli's exclusion principle. Now, plugging the Hamiltonian $H$ between the states $|\eta\rangle$ and $|\zeta\rangle$, and doing some 
mathematical steps we evaluate all the terms. For the three sub-Hamiltonians they are as follows.
\begin{eqnarray}
\langle \eta|H_1|\zeta\rangle &= & \epsilon_p \delta_{p,r} \delta_{q,s} + \epsilon_q \delta_{p,r} \delta_{q,s}, \nonumber \\
\langle \eta|H_2|\zeta\rangle &=& t_x e^{i\theta} \left(\delta_{p,r+1} \delta_{q,s}+\delta_{q,s+1} \delta_{p,r}\right) \nonumber \\
 &+& t_x e^{-i\theta} \left(\delta_{p+1,r} \delta_{q,s}+\delta_{q+1,s} \delta_{p,r}\right), \nonumber \\
\langle \eta|H_3|\zeta\rangle &= & U \delta_{p,q,r,s}. \nonumber
\end{eqnarray}
Using these factors we can easily compute all the matrix elements with the help of a simple algorithm to get the full many-body 
Hamiltonian matrix. Following the above steps of 1D ring with two opposite spin electrons, we can now construct the Hamiltonian
of any multi-ring system with higher number of up and down spin electrons. Finally, we find the many-body energy eigenvalues by 
diagonalizing the matrix. As already mentioned, one important limitation arises in this step which is the diagonalization of the full
matrix since the dimension of the matrix gets enhanced so rapidly with increasing the system size and the number of electrons. 
As illustrative example, for a $N_x$-site ring with two opposite spin electrons, the dimension is $N_x^2\times N_x^2$. Thus, for 
a nanotube having (say) $N_x=N_y=20$, the dimension of the matrix when it contains only these two opposite spin electrons becomes 
$400^2\times 400^2$, which is impossible to handle (specifically within our computational facilities). Thus, working with the full 
many-body Hamiltonian, we need to restrict ourselves within small size systems with few number of electrons.

\subsection{Hartree-Fock Mean-Field theory}

In the MF scheme~\cite{pc30,pc31,pc32,pc33,pc34} we can manage to find the energy eigenvalues of large systems. Here the fluctuations 
are neglected assuming they are small 
enough around the mean densities. This is the key concept of mean-field approximation (MFA). The MFA allows us to decompose the interacting 
Hamiltonian into two non-interacting parts. One is associated with the up spin electrons, while the other is involved with the down spin 
ones. Under MFA, the TB Hamiltonian Eq.~\ref{eqn1} is expressed as
\begin{eqnarray}
H_{MF} &=& H_{\uparrow} + H_{\downarrow} -U\sum_{j,k}\langle c_{j,k,\uparrow}^{\dagger} c_{j,k,\uparrow} \rangle 
\langle c_{j,k,\downarrow}^{\dagger} c_{j,k,\downarrow} \rangle \nonumber \\
 &=& H_{\uparrow} + H_{\downarrow} -U\sum_{j,k}\langle n_{j,k,\uparrow} \rangle \langle n_{j,k,\downarrow} \rangle
\label{eqn3}
\end{eqnarray}
where $n_{j,k,\sigma}$ ($=c_{j,k,\sigma}^{\dagger}c_{j,k,\sigma}$) corresponds to the number operator. 
$\langle n_{j,k,\sigma} \rangle$ is the expectation value of the number operator, which measures the mean occupation of electrons 
i.e., the electron density at the site ($j,k$) with spin $\sigma$. The Hamiltonians $H_{\uparrow}$ and $H_{\downarrow}$ read as
\begin{eqnarray}
H_{\uparrow}&=& t_y \sum_{j,k} c_{j,k,\uparrow}^{\dagger} c_{j+1,k,\uparrow} +t_x \sum_{j,k} e^{i\theta} c_{j,k,\uparrow}^{\dagger} 
c_{j,k+1,\uparrow} + h.c. \nonumber \\
&+& \sum_{j,k} \left(\epsilon_{j,k,\uparrow} + U \langle n_{j,k,\downarrow} \rangle\right) c_{j,k,\uparrow}^{\dagger} c_{j,k,\uparrow}
\label{eqn4}
\end{eqnarray}
and 
\begin{eqnarray}
H_{\downarrow}&=&t_y \sum_{j,k} c_{j,k,\downarrow}^{\dagger} c_{j+1,k,\downarrow} + t_x \sum_{j,k} e^{i\theta} c_{j,k,\downarrow}^{\dagger} 
c_{j,k+1,\downarrow} + h.c. \nonumber \\
&+& \sum_{j,k} \left(\epsilon_{j,k,\downarrow} + U \langle n_{j,k,\uparrow} \rangle\right) c_{j,k,\downarrow}^{\dagger} 
c_{j,k,\downarrow}.
\label{eqn5}
\end{eqnarray}
As $H_{\uparrow}$ and $H_{\downarrow}$ are single particle Hamiltonians, we can easily handle them. The last term in 
Eq.~\ref{eqn3} is simply a number, as it is the product of the averages of two number operators. 

The final 
solution i.e., the energy eigenvalues are obtained in a self-consistent procedure. We start with a set of initial guess values
of $\langle n_{j,k,\uparrow} \rangle$ and $\langle n_{j,k,\downarrow} \rangle$, and determine the energy eigenvalues and eigenvectors.
Using the eigenvectors a new set of $\langle n_{j,k,\uparrow} \rangle$ and $\langle n_{j,k,\downarrow} \rangle$ is formed, and then we 
compute the energy eigenvalues from $H_{\uparrow}$ and $H_{\downarrow}$. This process continues till the converged solution is reached.

\subsection{Calculation of ground state energy}

To calculate flux driven circular current we need to know the ground state energy of the system. It is obtained in two ways, for the 
two different prescriptions i.e., diagonalization of the full many-body Hamiltonian and the HF MF scheme. For the previous one, since 
we deal with the many-body states, the ground state energy $E_g$ at absolute zero temperature becomes the lowest energy eigenvalue of 
the many-body Hamiltonian. While, in the other prescription, $E_g$ is determined at zero temperature from the relation~\cite{pc30,gup1}
\begin{equation}
E_g=\sum_m E_{m,\uparrow} + \sum_m E_{m,\downarrow} - U \sum_{j,k}
\langle n_{j,k,\uparrow} \rangle \langle n_{j,k,\downarrow} \rangle
\label{eqn6}
\end{equation} 
where, $E_{m,\uparrow}$ and $E_{m,\downarrow}$ are the eigenenergies of the single particle Hamiltonians $H_{\uparrow}$ 
and $H_{\downarrow}$ (Eqs.~\ref{eqn4} and \ref{eqn5}), respectively. Depending on the filling factors associated with up and down spin
electrons ($N_{\uparrow}$ and $N_{\downarrow}$) we take the sum of the lowest $N_{\uparrow}$ and $N_{\downarrow}$ energy levels to get
the ground state energy $E_g$.

\subsection{Calculation of persistent current}

Finally, we evaluate the current in the nanotube, that appears due to the AB flux $\phi$, from the relation~\cite{pc2}
\begin{equation}
I=-\frac{\partial E_g}{\partial \phi}.
\end{equation}
It clearly suggests that the nature of flux driven current solely depends on the slope of the ground state energy with respect to the 
flux $\phi$. Vanishing slope yields zero circular current, otherwise a finite current appears.

\section{Numerical Results and Discussion}

In this section we present our numerical results those are computed by using the above theoretical framework. 
As the sites are non-magnetic in nature, we have $\epsilon_{j,k,\uparrow}=\epsilon_{j,k,\downarrow}$. For the ordered case we set them 
to zero without loss of any generality, while they are chosen `randomly'~\cite{pc18,pc19,pc22,kus} from a `Box' distribution function 
of width $W$. For the entire numerical calculations we select $t_x=t_y=1\,$eV, and measure the other energies in unit of $t_x$. 
All the circular currents are scaled by the factor $40\,\mu$A. This typical scale factor is associated with the 
term $e/h$ ($\sim 40\,\mu$A/eV) that appears in the evaluation of current.   
\begin{figure*}[ht]
{\centering \resizebox*{6.5cm}{13cm}{\includegraphics{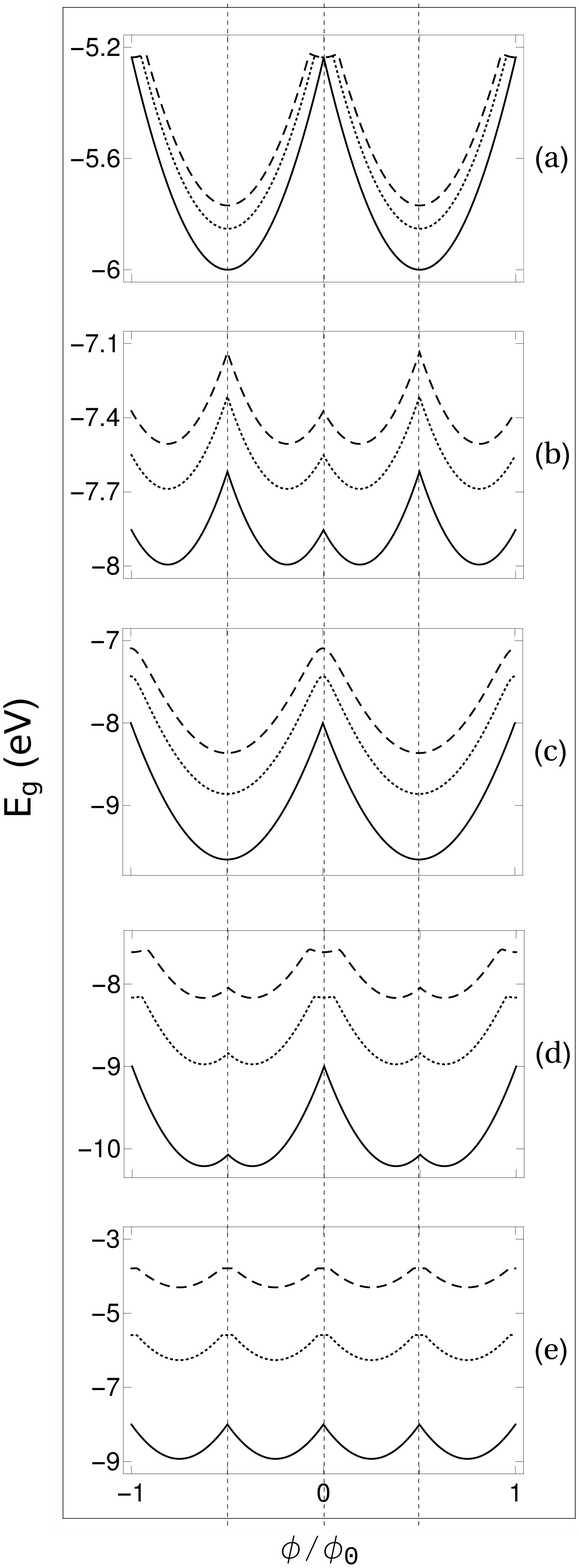}}
\centering \resizebox*{6.5cm}{13cm}{\includegraphics{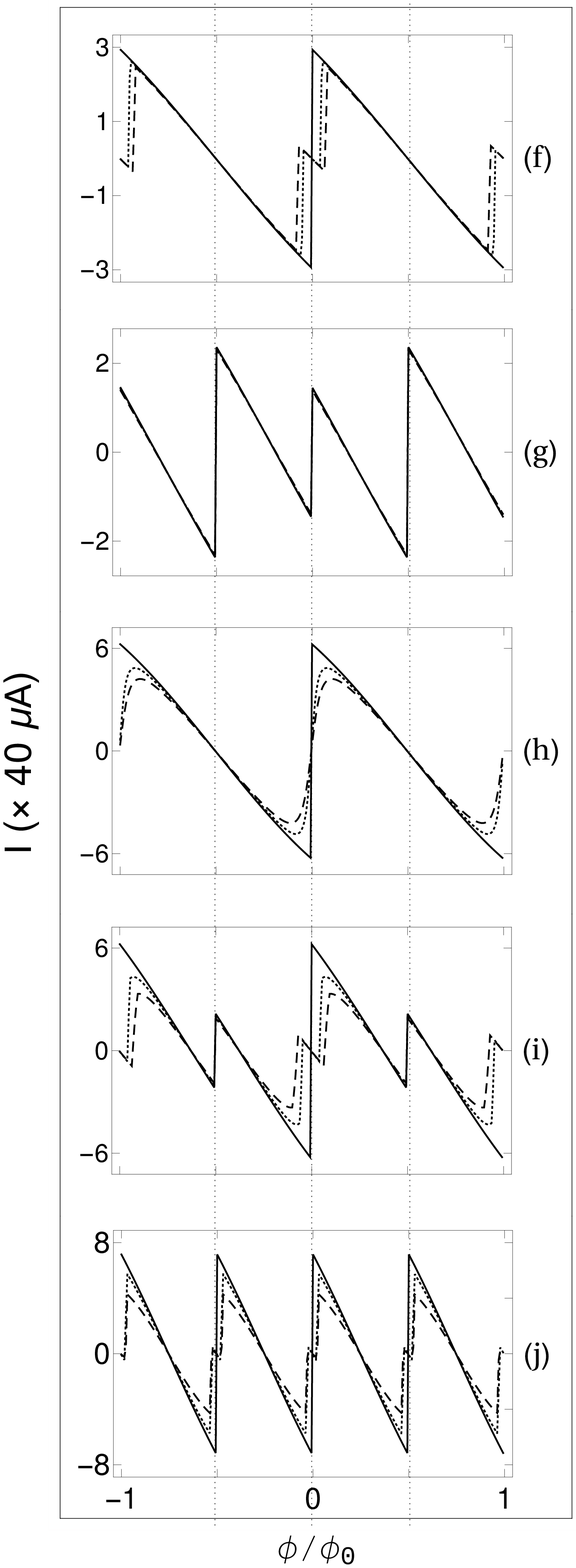}}\par}
\caption{Ground state energy and corresponding current as a function of flux $\phi$ for ordered multi-ring systems with different 
combinations of up and down spin electrons. In each spectrum, the results are shown for three different values of $U$, where the solid, 
dotted and dashed curves correspond to $U=0$, $2$ and $4$, respectively. The number of lattice sites in each ring and the combination 
of up and down spin electrons in different cases are as follows: (a) and (f) $N_x=5$, $N_{\uparrow}=N_{\downarrow}=1$; (b) and (g)
$N_x=5$, $N_{\uparrow}=2$, $N_{\downarrow}=1$; (c) and (h) $N_x=4$, $N_{\uparrow}=N_{\downarrow}=2$; (d) and (i) $N_x=4$, $N_{\uparrow}=3$, 
$N_{\downarrow}=2$; (e) and (j) $N_x=3$, $N_{\uparrow}=N_{\downarrow}=3$. In all these cases we set $N_y=2$.} 
\label{fig2}
\end{figure*}

\subsection{Fully ordered multi-ring system}

Before focusing on the central parts of our analysis, we start with the energy-flux and current-flux characteristics for the completely 
ordered system to understand the precise role of e-e correlation on these quantities. The results are presented in Fig.~\ref{fig2}, 
where everything is done by considering the full many-body Hamiltonian. Here it is relevant to note that for a `multi-ring system', no 
analysis has been made so far using the full Hamiltonian. Five different cases are taken into account depending on the number of up and 
down spin electrons those are: ($\uparrow,\downarrow$), ($\uparrow,\uparrow,\downarrow$), ($\uparrow,\uparrow,\downarrow,\downarrow$),
($\uparrow,\uparrow,\uparrow,\downarrow,\downarrow$) and ($\uparrow,\uparrow,\uparrow,\downarrow,\downarrow,\downarrow$). In the left column 
ground state energies are shown (Figs.~\ref{fig2}(a)-(e)) for the three different values of $U$, and the corresponding currents are 
presented in the right column (Figs.~\ref{fig2}(f)-(j)).
Several atypical features are observed those are thoroughly discussed one by one as follows.

\vskip 0.2cm
\noindent
\underline{Case I. Two-electron system}: For the system with two electrons ($\uparrow,\downarrow$) a sharp change in 
slope in $E_g$ takes place across $\phi=0$ for $U=0$ (solid line of Fig.~\ref{fig2}(a)) which yields a saw-tooth like behavior in the 
current (solid line of Fig.~\ref{fig2}(f)). This is a well established fact~\cite{pc2}. But as we include $U$, a dip-like structure 
appears in the $E_g$-$\phi$ curve around $\phi=0$ (dotted line of Fig.~\ref{fig2}(a)), and most importantly, it arises for any non-zero 
value of $U$. The arc length of the dip increases with $U$ (dashed line of Fig.~\ref{fig2}(a)). Because of this finite arc in $E_g$-$\phi$ 
curve, kink appears in the current-flux characteristics as clearly seen from the dotted and dashed curves of Fig.~\ref{fig2}(f). The
origin of the kink like structure in current followed by the dip in ground state energy is clearly described below. Now, comparing 
the spectra given in Fig.~\ref{fig2}(f) we see that the current amplitude gets reduced with $U$. This is solely associated with
the mutual repulsion of up and down spin electrons which prohibits their motions. Due to the less number of electrons the reduction of
current is not so prominent, but it can be more properly seen from the other systems having higher number of electrons. Both the energy and
current show $\phi_0$ periodicity, which is commonly observed in almost all the earlier studies~\cite{pc1,pc2} involving circular current
in AB loops. 

\vskip 0.2cm
\noindent
{\em Origin of the dip in $E_g$, and hence, the kink in $I$ due to $U$}: Appearance of a dip in the ground state 
energy is an interesting phenomenon which yields a kink-like structure in the $I$-$\phi$ curve since the current is directly proportional
to the first order derivative of $E_g$ with respect to $\phi$. For the system with two opposite spin electrons we can write the full 
many-body 
Hamiltonian matrix in a `block diagonal' form by selectively choosing the basis states from two different sub-spaces associated with net 
spin $S$. Here we have two possibilities of $S$: $S=0$ and $S=1$. For instance, a basis state like 
$(1/\sqrt{2})(|\uparrow,\downarrow,0,0,...\rangle-|\downarrow,\uparrow,0,0,...\rangle)$ belongs to the spin sub-space with $S=0$, while a 
state like $(1/\sqrt{2})(|\uparrow,\downarrow,0,0,...\rangle+|\downarrow,\uparrow,0,0,...\rangle)$ lies in the other spin sub-space i.e., 
$S=1$. The eigenstates of the block matrix of the spin sub-space $S=1$ are no longer affected by the Hubbard correlation $U$ which we 
confirm through our exhaustive numerical calculations. Only the eigenstates of the other block matrix for the spin sub-space $S=0$ are 
influenced by the electronic correlation. When $U=0$, the $U$-independent and $U$-dependent energy levels exactly touch each other at 
$\phi=0$. On the other hand, for finite $U$ the $U$-independent energy level remain unchanged whereas the energies of the other levels 
get increased due to repulsive interaction. Because of this, the $U$-independent energy level becomes the lowest energy level within a 
small window across $\phi=0$ which provides a dip in $E_g$. As the energies of the $U$-dependent levels increase, we get more higher 
dip with $U$. The dip in $E_g$-$\phi$ curve results a kink in the current spectrum.

\vskip 0.2cm
\noindent
\underline{Case II. Three-electron system}: For the system with three electrons ($\uparrow,\uparrow,\downarrow$), 
the slope of the energy and the current are very less sensitive even at moderate $U$ (Figs.~\ref{fig2}(b) and (g)). Unlike the 
two-electron system, 
here a large kink appears around $\phi=0$ even when $U=0$. The appearance of the kink is solely related to the variation of $E_g$ which 
on the other hand depends on the number of electrons in a system. The less sensitivity of current on $U$ for this typical three-electron
case can be physically explained as follows. As the system contains two up and one down spin electrons, mutual repulsion between one up 
and one down spin electrons occurs while there is always another up spin electron that can move in the geometry without facing the repulsive 
effect. Thus, for the low or even moderate $U$, the effect of $U$ in the current spectrum is not properly seen. For large enough $U$, minor 
change is obtained which we confirm through our detailed numerical calculations. The current provides the usual $\phi_0$ periodicity. Here 
it is worthy to note that, instead of considering a two-ring system with $10$ lattice sites (i.e., what is taken in our present case), if 
we would take a single ring ($N_y=1$) with $N_x=3$ then the current would exhibit $\phi_0/2$ periodicity. More precisely we can say that, 
the current shows $\phi_0/2$ periodicity for the odd half-filled case (not shown here in Fig.~\ref{fig2}(g)). Its origin will be understood 
from our forthcoming analysis. 

\vskip 0.2cm
\noindent
\underline{Case III. Four-electron system}: In the absence of electronic correlation ($U=0$), current shows a sharp transition at $\phi=0$
(solid line of Fig.~\ref{fig2}(h)) following the $E_g$-$\phi$ curve (solid curve of Fig.~\ref{fig2}(c)), like what is observed for the 
two-electron system. This phenomenon disappears when the Hubbard correlation is included. The sharp discontinuity in $I$-$\phi$ curve
is directly related to the sudden change in slope of $E_g$ with $\phi$. The later one is again involved with the crossing of the 
neighboring energy levels. For $U=0$, we find a crossing of the lowest energy level with the neighboring levels across $\phi=0$ which 
gives rise to a sudden change in slope. This is a general feature i.e., whenever there is an intersection among the two or more energy 
levels, a sudden change in slope occurs. On the other hand, for finite $U$ the energy levels are well separated without having any 
intersection. Accordingly we get a smooth variation. The usual reduction of
current with $U$ is due to the repulsive interaction between the opposite electrons. The reduction becomes more 
effective when the unoccupied lattice sites are less, and, eventually a situation can be made when all the sites are occupied by at least 
one electron (viz, the half-filled band case), then no electron can move from one site to the other. Under this condition, the system 
reaches to the insulating phase. Here also we get $\phi_0$ periodic circular currents.

\vskip 0.2cm
\noindent
\underline{Case IV. Five-electron system}: The results of the five-electron system ($\uparrow,\uparrow,\uparrow,\downarrow,\downarrow$) 
are shown in Figs.~\ref{fig2}(d) and (i).
The kink-like structure at different $\phi$ values is clearly understood from the variation of $E_g$. For this system the reduction of 
current with $U$ is quite prominent since the number of electrons is quite higher compared to the previous cases. Among the five 
electrons, there are two pairs of up and down spin electrons, and hence the repulsive interaction dominates over the motion of other 
single electron that can get unoccupied sites to hop. For this five-electron system, the energy and the current exhibit conventional 
$\phi_0$ periodicity. 

\vskip 0.2cm
\noindent
\underline{Case V. Six-electron system}: Finally, we focus on the spectra Figs.~\ref{fig2}(e) and (j) where the results of the 
six-electron ($\uparrow,\uparrow,\uparrow,\downarrow,\downarrow,\downarrow$) system are presented. 
For this case the system is half filled as six electrons are accommodated among the six sites. Both at $\phi=0$ and $\phi=\pm 0.5$, 
kinks appear in the current following the energy-flux characteristics, as long as the e-e interaction is included. This is similar to 
what we get in the other electron systems studied above. But, the notable observation is that, unlike $\phi_0$ 
periodicity, here we get $\phi_0/2$ periodicity. This is quite atypical compared to the reported~\cite{pc6,pc28} $\phi_0/2$ periodicity 
for the `ensemble averaged' current, as in our case no such averaging is done. The appearance of $\phi_0/2$ periodicity is one of the 
key features of our analysis, and below we make an attempt to explain this fact with a suitable mathematical description. Maybe more
comprehensive and simpler explanations will be available in near future.

\vskip 0.2cm
\noindent
{\bf Possible origin of getting $\phi_0/2$ periodic current -- An analytical treatment}:
To understand the basic mechanism, let us start with a simple system, as a typical example, which is a one-dimensional five-site 
($N_x=5$) non-interacting ($U=0$) AB ring with five electrons ($\uparrow,\uparrow,\uparrow,\downarrow,\downarrow$). For this perfect ring, 
we set the site energy to zero, without loss of any generality. Since the periodicity in current is directly linked with the
variation of ground state energy with $\phi$, we concentrate on the ground state energy $E_g(\phi)$.

Suppose at any arbitrary flux $\phi_1$, the five eigenvalues of the ring (arranged in higher order) are: 
$-a$, $-b$, $-c$, $d$ and $e$, and they should obey the relation $-a-b-c+d+e=0$ i.e.,
\begin{equation}
d+e=a+b+c
\label{apeq1}
\end{equation}
since the sum of the eigenvalues is equal to the trace of the Hamiltonian matrix of the ring. Now, we consider another magnetic
flux $\phi_2$ ($\phi_1\ne\phi_2$) and imagine that for $\phi=\phi_2$ all the above mentioned eigenvalues get flipped with a negative sign
i.e., they are $-e$, $-d$, $c$, $b$ and $a$. Thus, they also satisfy the relation
\begin{equation}
a+b+c=d+e.
\label{apeq2}
\end{equation}
Hence, for the half-filled band case the ground state energy at $\phi_1$ is $E_g(\phi_1)=-2a-2b+c$, and at $\phi_2$ it becomes
$E_g(\phi_2)=-2e-2d+c$. Using Eq.~\ref{apeq1}, we see that
\begin{equation}
E_g(\phi_1)=E_g(\phi_2).
\label{apeq3}
\end{equation}
For even $N_x$, reaching to the above condition is quite obvious if similar swapping of energy levels takes place at these two distinct
fluxes.

Thus, swapping the energy eigenvalues is the necessary condition to get identical ground state energy at two different AB fluxes. Now,
the question is that, {\em if we set $\phi_1$ at any particular value within one flux-quantum window, then can we find any other flux 
$\phi_2$ within this window such that $E_g(\phi_1)=E_g(\phi_2)$}.

It can be answered with the help of the energy eigenvalue equation for a $N_x$-site AB ring which is
\begin{equation}
E_n(\phi)=2 t_x \cos\left[\frac{2\pi}{N_x}\left(n+\frac{\phi}{\phi_0}\right)\right]
\label{apeq4}
\end{equation}
where $n$ is the energy level index. Thus, for any two energy levels, (say) $n_1$ and $n_2$, the eigenvalues at $\phi_1$ and $\phi_2$
are:
\begin{eqnarray}
E_{n_1}(\phi_1) & = & 2 t_x \cos\left[\frac{2\pi}{N_x}\left({n_1}+\frac{\phi_1}{\phi_0}\right)\right],
\label{apeq5} \\
E_{n_2}(\phi_2) & = & 2 t_x \cos\left[\frac{2\pi}{N_x}\left({n_2}+\frac{\phi_2}{\phi_0}\right)\right].
\label{apeq6}
\end{eqnarray}
Now, to have $E_{n_1}(\phi_1)=-E_{n_2}(\phi_2)$, we get the condition from Eq.~\ref{apeq5} and Eq.~\ref{apeq6} as
\begin{equation}
\frac{2\pi}{N_x}\left({n_2}+\frac{\phi_2}{\phi_0}\right)-\frac{2\pi}{N_x}\left({n_1}+\frac{\phi_1}{\phi_0}\right)=\pi
\label{apeq7}
\end{equation}
which simplifies to
\begin{equation}
(n_2-n_1)+\frac{(\phi_2-\phi_1)}{\phi_0}=\frac{N_x}{2}.
\label{apeq8}
\end{equation}
From Eq.~\ref{apeq8}, we can now clearly explain the origin of $\phi_0/2$ periodicity. As ($n_2-n_1$) is also an
integer, $\phi_2-\phi_1$ can achieve the value $\pm \phi_0/2$ only when $N_x$ is `odd'. Thus, if we set $\phi_1=0$ then
perfect swapping takes place at $\phi_2=\pm \phi_0/2$ which results the half flux-quantum periodicity. On the other hand, for even
$N_x$, $\phi_2-\phi_1$ becomes identical to $\phi_0$, which yields the conventional $\phi_0$ flux-quantum periodicity.

The above mathematical analysis can easily be extended in multi-ring systems to justify the occurrence of $\phi_0/2$ periodicity. 
From the physical point of view it is inferred that, {\em the central mechanism of getting such half flux-quantum periodicity for 
the odd half-filled case relies on the particle-hole symmetry of the system.}

\subsection{Interplay between e-e interaction and disorder: Fully disordered and ODS nanotubes}

\begin{figure}[ht]
{\centering \resizebox*{8.25cm}{5.5cm}{\includegraphics{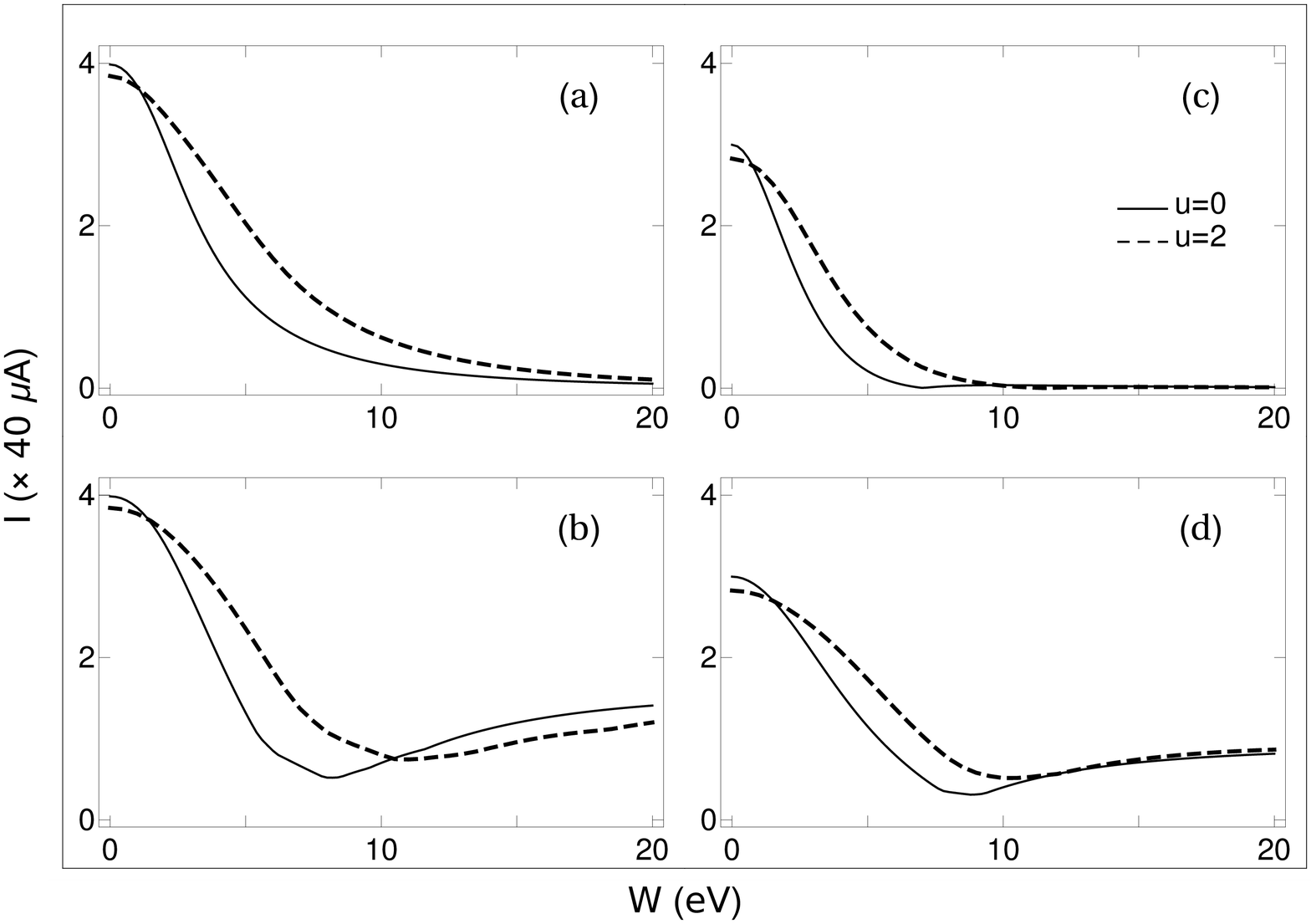}}\par}
\caption{Absolute current, computed at a typical magnetic flux $\phi=0.2$, as a function of the impurity strength $W$ for the FD (shown 
in (a) and (c)) and ODS (given in (b) and (d)) systems. Here we take $N_y=2$ and $N_x=4$. The left and the right columns represent the 
systems with four ($\uparrow,\uparrow,\downarrow,\downarrow$) and five ($\uparrow,\uparrow,\uparrow,\downarrow,\downarrow$) electrons, 
respectively. The results are worked out for both the non-interacting and interacting cases.}
\label{fig3}
\end{figure}

The situation becomes more interesting and complicated as well, when we include impurities along with the Hubbard correlation.
Two different cases are considered depending on the distribution of impurities in the system, and they are categorized 
as (i) full disordered 
(FD) and (ii) ordered-disordered separated (ODS) ones. For a FD nanotube, impurities are added at all the lattice sites as conventionally
used, whereas for the ODS nanotube, impurities are given in one half of the tube keeping the other part free. In Fig.~\ref{fig3} we 
present the variations of current 
amplitudes, computed at a typical magnetic flux $\phi=0.2$, by changing the impurity strength $W$ in a wide range. The results of $U=0$ 
are also superimposed in the current spectra to have a clear comparison between the interacting and non-interacting ($U\ne 0$) cases. 
A sharp contrasting behavior is obtained for these two kinds of disordered multi-ring systems. In the presence of 
uniform disorder, current amplitude gradually decreases with $W$ and eventually drops to zero, irrespective of the e-e interaction 
strength (Figs.~\ref{fig3}(a) and (c)). This reduction is fully associated with the impurity induced electronic localization which
is known as Anderson localization~\cite{pc18,pc19,anlc1,anlc2,anlc3,anlc4}. More disorder yields more localization, and in the limit 
of large $W$ current amplitude practically vanishes, as expected. Comparing the current spectra, it is clearly seen that the rate of fall 
of current with $W$ becomes slower for the interacting system than the non-interacting one. This is because of the repulsive 
nature of electrons which does not favor to sit two opposite spin electrons in one site. For the disordered system, the repulsive
Coulomb interaction eventually leads to a delocalization and thus enhanced current is obtained compared to the interacting free one.  

The effect of $W$ is reasonably different for the ODS system, where a gradual reduction of current is not seen over the full window, 
rather a completely 
opposite signature is obtained beyond a critical value of $W$ (Figs.~\ref{fig3}(b) and (d)). In the limit of weak $W$, current 
gets reduced with $W$ like what we get in the FD system. Reaching to a minimum at moderate $W$, the current starts to enhance with
the impurity strength and it achieves a sufficiently large value when $W$ is too high. The underlying physical mechanism for this 
atypical nature is described as follows. The ODS system can be considered as a coupled system where the ordered region is directly 
coupled to the disordered one. The coupling between these two regions gradually reduces with $W$, and eventually drops almost to zero
when $W$ is too large. Thus, for the weak $W$ as the disordered region is coupled to the ordered part, the energy eigenstates gets 
affected by the impurities which results a reduction of current with increasing the impurity strength. At the critical limit the 
scattering effect becomes maximum resulting a minimum current amplitude. Beyond that limit, the states start to get less influenced by the 
impurities since then the decoupling between the two regions becomes more prominent. Finally, when $W$ is too large, the ordered and
disordered parts are practically decoupled, and then the contribution towards current comes from the ordered region. 
\begin{figure}[ht]
{\centering \resizebox*{8cm}{3.25cm}{\includegraphics{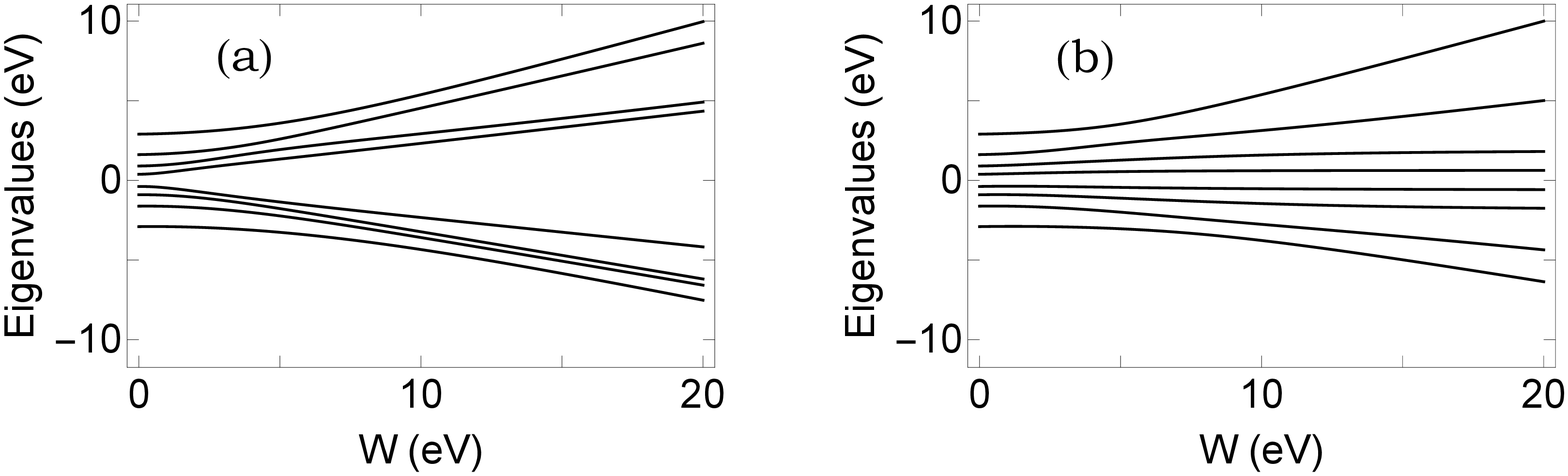}}\par}
\caption{Energy eigenvalues as a function of the impurity strength $W$ for the two-ring ($N_y=2$) non-interacting ($U=0$) system, where 
each of these rings contains $4$ lattice sites. Here (a) and (b) correspond to the FD and ODS systems, respectively. For the ODS case, 
the decoupling between the ordered and disordered regions in the limit of high $W$ can be easily followed.}
\label{fig33}
\end{figure}
The delocalization
process gets enhanced further with the inclusion of Hubbard correlation, and accordingly, the critical point of current minimum gets 
shifted to a larger $W$. Both for the zero and non-zero values of $U$, the signature of the current with impurity strength remains same,
as clearly seen from the spectra. In the weak disorder regime the many-body ground state becomes the `coherent'
linear combination of the localized states, emerged from the disordered zone, and the delocalized states that are generated from the 
ordered counterpart. On the other hand, in the strong disorder regime, the ground state becomes an `incoherent' product of these states.
During the transition from weak-to-strong disorder region, current provides a soft minimum.
Now, to visualize the decoupling between the two different regions more clearly, in Fig.~\ref{fig33} 
we show the dependence of energy eigenvalues with $W$, by varying it in a wide range. For the sake of simplification, here we set $U=0$
so that single particle energy eigenvalues are obtained quite easily by diagonalizing the TB Hamiltonian. For the FD system, all the
eigenvalues are affected whether $W$ is small or large (see Fig.~\ref{fig33}(a)). On the other hand, among the eight energy 
levels of the ODS system, four are affected and the remaining four levels are almost undisturbed (Fig.~\ref{fig33}(b)), and it becomes 
more prominent when $W$ is quite large. It clearly suggests that the ordered region gets separated from its disordered counter part.

\begin{figure}[ht]
{\centering \resizebox*{6.5cm}{4.0cm}{\includegraphics{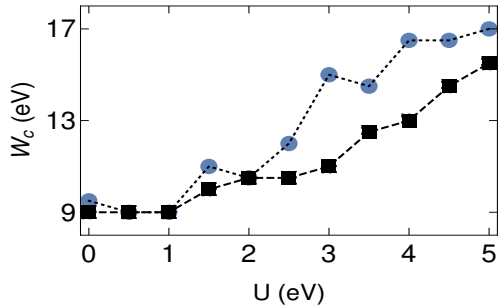}}\par}
\caption{Dependence of $W_c$ as a function of $U$ for the ODS system. The lines with filled black boxes and blue circles are associated 
with the four and five-electron cases as taken in Fig.~\ref{fig3}, respectively. The system size and the other physical parameters remain 
also same as considered in Fig.~\ref{fig3}. The results are worked out by averaging over $100$ distinct disordered configurations.}
\label{fig4}
\end{figure}

From the results of the ODS system, presented in Figs.~\ref{fig3}(b) and (d), a question naturally comes that how does $W_c$ (critical 
$W$ for which the current achieves a minimum) vary with the interaction strength $U$. The results are shown in Fig.~\ref{fig4} and they 
are computed for the identical electron systems as taken in Fig.~\ref{fig3}. For each $U$, the critical $W$ is obtained by scanning the
current amplitude over a wide range of $W$. Both for the four- and five-electron cases, $W_c$ shows an increasing behavior 
providing almost a linear dependence with $U$. For other set of electrons, we also get a similar kind of variation which we confirm 
through our exhaustive numerical calculations. This enhancement of $W_c$ is solely associated with the electronic delocalization in 
the presence of e-e correlation, as already stated above.
From this $W_c$-$U$ diagram we can estimate the high and low conducting natures depending on the other physical parameters of the system.
 
\subsection{Mean-field results}

All the characteristic features analyzed so far are worked out by diagonalizing the full many-body Hamiltonian, and therefore the results
are restricted in small size systems. To examine whether the features studied above still persist even for larger ring sizes as well, we 
now focus on the mean-field results those are computed for much larger systems than the previous ones. Before that, first we need to 
check the
\begin{figure}[ht]
{\centering \resizebox*{6.5cm}{4.0cm}{\includegraphics{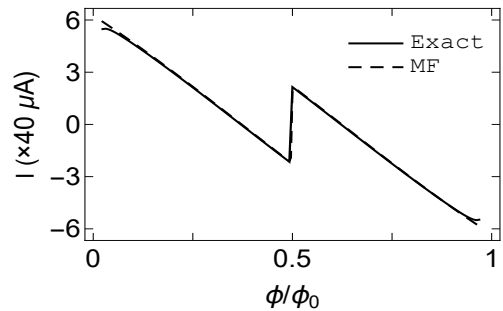}}\par}
\caption{Comparison between the results obtained from the exact diagonalization of the full many-body Hamiltonian and the HF MF approach. 
Currents are calculated for the ODS two-ring system having five electrons ($\uparrow,\uparrow,\uparrow,\downarrow,\downarrow$), where each 
ring contains four sites. The other physical parameters are: $U=0.5$ and $W=2$. Averaging over same number of distinct disordered
configurations is taken into account, as considered in Fig.~\ref{fig4}.}
\label{fig5}
\end{figure}
accuracy and reliability of the MF results, particularly in the presence of random impurities. For that, in Fig.~\ref{fig5} we make a 
comparison between the currents obtained in the two methods used in our work. As the result of the full TB many-body 
Hamiltonian is 
shown along with the MF result, here we choose a small size system (eight lattice sites) with five electrons (three up and two down). 
The currents evaluated in these two separate prescriptions almost overlap with each other for the entire flux-quantum window. Thus, we 
can safely use MF results for further analysis. 
Here it is relevant to note that finding an optimized converged solution in the MF scheme 
is the most important issue. In the absence of impurities, one can get the converged solution quite easily even for large value of $U$.
But as long as impurities are given, it takes much longer time to reach into the final solution, and the situation becomes more 
tedious for higher values of $U$. In that case we need to check the results with different initial guess values 
$\langle n_{j,k,\uparrow}\rangle$ and $\langle n_{j,k,\downarrow}\rangle$, such that no spurious result appears.

\begin{figure}[ht]
{\centering \resizebox*{6.5cm}{4.0cm}{\includegraphics{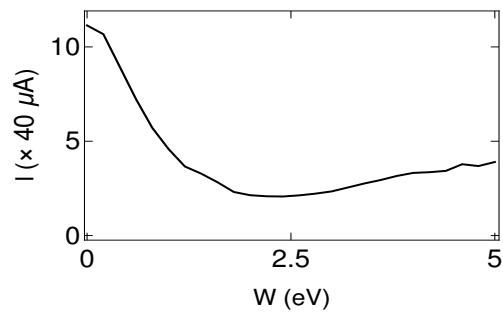}}\par}
\caption{Absolute current, computed at a typical magnetic flux $\phi=\phi_0/4$, as a function of $W$ for a much bigger ODS nanotube having
$N_x=10$ and $N_y=8$. Here we set $U=0.5$ and choose $N_{\uparrow}=N_{\downarrow}=35$.}
\label{fig6}
\end{figure}

Considering a much bigger ODS nanotube, in Fig.~\ref{fig6} we show the variation of absolute current as a function of the disorder strength
$W$. The nature of the current spectrum remains unchanged like what we get for smaller rings (Figs.~\ref{fig3}(b) and (d)). In the limit
of weak $W$, current gets reduced with $W$ providing a localizing behavior, while the delocalization process gets started when $W$
crosses its critical value. This is the generic feature of an ODS nanotube and can be exploited in different ways. In 
addition, the 
other notable aspect is that once the system goes to the delocalized phase, it never comes back to the localized phase even at too 
large $W$ since for such a system {\em mobility edge} (ME) always persists. This behavior is completely different compared to the uniform 
random disordered 3D lattices, where ME vanishes at large disorder. The ME separates the localized states from the conducting 
ones~\cite{skmprl}, which thus yields a localization-to-delocalization transition. In analyzing the behavior of $I$-$W$ characteristics 
of the ODS system (Figs.~\ref{fig3}(b) and (d)), it is explained that the coupling between the ordered and disordered sectors gradually 
decreases with $W$, and for large enough $W$ these two regions are practically decoupled (the phenomenon can also be understood clearly 
from Fig.~\ref{fig33}(b)). Under this situation, localized states are emerged from the disordered part, while extended states are generated 
from the other region. Therefore, a mixture of both the localized and extended states are obtained for the ODS system which leads to the 
existence of a ME. On the other hand, for the traditional `uniform'
\begin{figure}[ht]
{\centering \resizebox*{8.5cm}{9cm}{\includegraphics{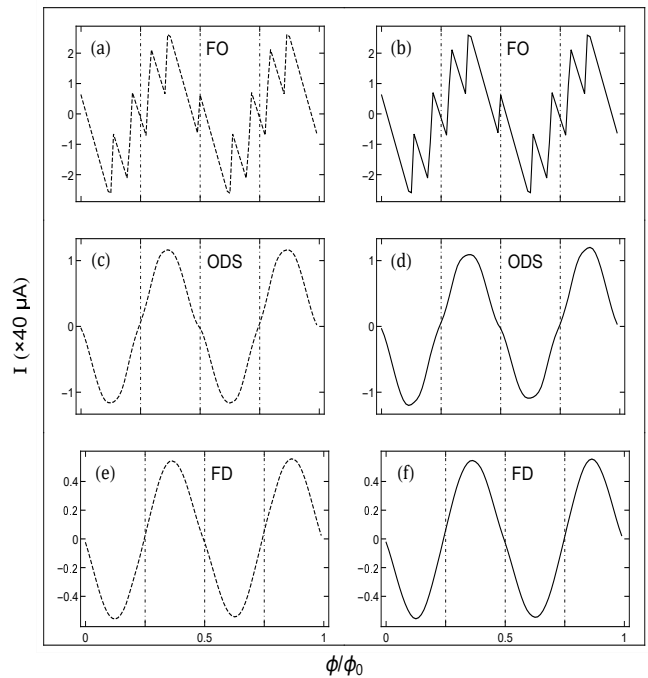}}\par}
\caption{Current-flux characteristics for `odd half-filled' nanotubes considering $N_x=7$ and $N_y=5$, where (a) and (b) correspond to
the fully ordered (FO), (c) and (d) denote ODS, and (e) and (f) represent FD nanotubes, respectively. Here we set $W=2$ and choose 
$N_{\uparrow}=18$ and $N_{\downarrow}=17$. The dashed and solid curves are associated with $U=0$ and $U=0.02$, respectively. For the 
ODS case, impurities are included in the upper two rings of the tube, keeping the other three rings free. In presence of disorder, 
averaging over $2000$ distinct random configurations are taken into account for each case.}
\label{fig7}
\end{figure}
random disordered 3D lattices `all the states' become localized for large $W$, as in this case there is no ordered region that can be
decoupled from the disordered one. Hence, for a uniform disordered lattice ME does not exist, and one cannot get the conducting phase 
at large $W$, unlike an ODS system.

Finally, we focus on the current spectra shown in Fig.~\ref{fig7}, where our main intentions are to check whether (i)
we can get half flux-quantum periodicity in the odd half-filled band case in a multi-channel system i.e., in a  nanotube and (ii) this 
behavior still persists even in the presence of disorder. To substantiate these facts, we present the results of three different kinds
of nanotubes viz, fully ordered (FO), ordered-disordered separated (ODS) and fully disordered (FD), and in each nanotube we choose
$N_x=7$ and $N_y=5$, with $N_{\uparrow}=18$ and $N_{\downarrow}=17$ (half-filled band case). In the disordered systems (ODS and FO),
we compute the currents taking the average over a large number ($2000$) of distinct random configurations such that more accurate results
are obtained. For the ordered nanotube, current exhibits half flux-quantum periodicity both in the absence and presence of e-e correlation 
(Figs.~\ref{fig7}(a) and (b)), and this feature is expected from our previous analysis. But the interesting fact is that, even for the 
disordered nanotube, be it fully or partly disordered, the phenomenon of half flux-quantum periodicity still persists which is clearly 
reflected from the spectra given in Figs.~\ref{fig7}(c)-(f). From our detailed numerics we find that, the $\phi_0/2$ periodicity is
always obtained, independent of the choice of the number of rings in the nanotube, as long as $N_x$ is `odd' and the system is half-filled.
This is a direct consequence of particle-hole symmetry. Once we set even $N_x$, no such phenomenon is observed and we get back to the 
conventional $\phi_0$ periodicity.

\section{Experimental Feasibility} 

Nowadays the designing of ring conductors at nanoscale level is not at all complicated due to the availability of 
advanced nanotechnologies. Several approaches are there like silicon etching process~\cite{fb1}, electron beam lithography~\cite{pc13}, 
nanosphere lithography~\cite{fb3,fb4}, and to name a few. With the help of these techniques different sized thin and thick rings, nanotubes 
and other tailor made geometries can be fabricated. Very recently Pham {\em et al.} have established a new approach of devising a ring 
like structure on a semiconducting surface using the atom manipulation technique~\cite{pham}. With this prescription atomically precise 
quantum ring can be designed. In another work Keyser {\em et al.} performed an experiment almost two decades ago to explore the influence 
of Coulomb correlation on Aharonov-Bohm oscillations considering a nano ring with less than ten electrons~\cite{keys}. Usually Coulomb 
interaction becomes significant in smaller rings with few number of electrons since in this case the interaction is not screened so much. 
All the physical parameters chosen in our theoretical work are consistent with realistic systems, and thus, we strongly believe that the 
results studied here can be tested in a suitable laboratory.
   
\section{Closing Remarks}

To conclude, in the present work we investigate theoretically magnetic response of interacting electrons in a `spatially non-uniform
disordered system'. The response is measured by the flux driven circular current in a multi-channel nanotube whose one segment is subjected
to disorder keeping the other part free. A sharp contrasting 
behavior is obtained compared to the conventional uniform disordered material. The e-e correlation is included in the well known Hubbard 
form. Describing the quantum system using a tight-binding framework, we compute the results in two ways: (i) exactly diagonalizing the full
many-body Hamiltonian and (ii) Hartree-Fock mean-field scheme. The latter is used for bigger systems having higher number of electrons. 
Several important features emerge from our analysis those are summarized as follows.\\
$\bullet$ Because of the e-e interaction kink-like structure appears in the current following a sudden change in slope around $\phi=0$
and/or $\phi=\pm \phi_0/2$ in the ground state energy. These kinks are specifically associated with the correlation among the up and 
down spin electrons. \\
$\bullet$ For the odd half-filled band case, current always exhibits $\phi_0/2$ periodicity irrespective of the number of rings in the 
nanotube and disorderness. \\
$\bullet$ The results are worked out for different sizes of the nanotubes, and all the physical pictures are equally valid in all these 
cases as long as we restrict ourselves in the meso-scale region. In the asymptotic limit ($N\rightarrow \infty$), no current will appear 
since it is a mesoscopic phenomenon. \\
$\bullet$ Before discussing MF results, we test the accuracy of MF calculations and see that almost for the entire flux window the current
obtained in the MF scheme matches very well with the other method where many-body energy levels are determined. \\
$\bullet$ For the FD system, current amplitude decreases and eventually drops to zero with $W$ where all the states are localized because
of the disorder. On the other hand, a completely different scenario is obtained for the ODS nanotube, where delocalizing behavior is
obtained beyond a critical $W$. This is a robust phenomenon and does not vanish even in the large limit of $W$. \\
$\bullet$ Finally, the interplay between e-e correlation and disorder is also critically examined. Depending on the winning factor current
gets reduced or enhanced.
 
Before an end, we would like to note that working with many-body systems has always been a challenging task. Here we attempt to make 
an in-depth analysis to discuss some important phenomena of the interacting electrons in a unconventional disordered system, that might
be useful to analyze electronic and magnetic properties of several such other systems.

\section*{ACKNOWLEDGMENT}

SKM acknowledges the financial support of DST-SERB, Government of India, under the Project Grant Number EMR/2017/000504 for carrying 
out his research.


\begin{thebibliography}{99}

\bibitem{ab1} W. Ehrenberg and R. E. Siday, Proc. Phys. Soc. B
\textbf{62}, 8 (1949).

\bibitem{ab2} Y. Aharonov and D. Bohm, Phys. Rev. \textbf{115}, 
485 (1959).

\bibitem{pc1} M. B\"{u}ttiker, Y. Imry, and R. Landauer, Phys. Lett. A \textbf{96}, 365 (1983)

\bibitem{pc2} H. F. Cheung, Y. Gefen, E. K. Reidel, and W. H. Shih, Phys. Rev. B \textbf{37}, 6050 (1988).

\bibitem{pc3} R. A. Smith and V. Ambegaokar, Europhys. Lett. \textbf{20}, 161 (1992).

\bibitem{pc4} H. Bouchiat and G. Montambaux, J. Phys. (Paris) \textbf{60}, 2695 (1989).

\bibitem{pc5} S. K. Maiti, M. Dey, S. Sil, A. Chakrabarti, and S. N. Karmakar, Europhys. Lett. \textbf{95}, 57008 (2011).

\bibitem{pc6} V. Ambegaokar and U. Eckern, Phys. Rev. Lett. \textbf{65}, 381 (1990).

\bibitem{pc7} U. Eckern and A. Schmid, Europhys. Lett. \textbf{18}, 457 (1992).

\bibitem{pc8} H. F. Cheung and E. K. Riedel, Phys. Rev. Lett. \textbf{62}, 587 (1989).

\bibitem{pc9} S. K. Maiti, J. Chowdhury, and S. N. Karmakar, Phys. Lett. A \textbf{332}, 497 (2004).

\bibitem{pc10} P. A. Orellana and M. Pacheco, Phys. Rev. B \textbf{71}, 235330 (2005).

\bibitem{pc11} S. K. Maiti, J. Chowdhury, and S. N. Karmakar, J. Phys.: Condens. Matter \textbf{18}, 5349 (2006).

\bibitem{pc12} L. P. Levy, G. Dolan, J. Dunsmuir, and H. Bouchiat, Phys. Rev. Lett. \textbf{64}, 2074 (1990).

\bibitem{pc13} E. M. Q. Jariwala, P. Mohanty, M. B. Ketchen, and R. A. Webb, Phys. Rev. Lett. \textbf{86}, 1594 (2001).

\bibitem{pc14} V. Chandrasekhar, R. A. Webb, M. J. Brady, M. B. Ketchen, W. J. Gallagher, and A. Kleinsasser, Phys. Rev. Lett.
\textbf{67}, 3578 (1991).

\bibitem{pc15} H. Bluhm, N. C. Koshnick, J. A. Bert, M. E. Huber, and K. A. Moler, Phys. Rev. Lett. \textbf{102}, 136802 (2009).

\bibitem{pc16} M. Banerjee, B. Mal, and S. K. Maiti, Physica E \textbf{106}, 312 (2019). 

\bibitem{pc17} N. O. Birge, Science \textbf{326}, 244 (2009).

\bibitem{liang} S.-D. Liang, Z. D. Wang, and J.-X. Zhu, Solid State Commun. \textbf{98}, 909 (1996).

\bibitem{chen} H. Chen, E. Zhang, K. Zhang, S. Zhang, RSC Adv. \textbf{5}, 45551 (2015).

\bibitem{pc18} P. W. Anderson, Phys. Rev. \textbf{109}, 1492 (1958).

\bibitem{pc19} P. A. Lee and T. V. Ramakrsihnan, Rev. Mod. Phys. \textbf{57}, 287 (1985) and references therein.

\bibitem{pc20} G. Bergmann, Eur. Phys. J. B \textbf{92}, 79 (2019).

\bibitem{pc21} R. Kotlyar, C. A. Stafford, and S. Das Sarma, Phys. Rev. B \textbf{58}, 3989 (1998).

\bibitem{pc22} M. Abraham and R. Berkovits, Phys. Rev. Lett. \textbf{70}, 1509 (1993). 

\bibitem{pc23} E. Gambetti-C\'{e}sare, D. Weinmann, R. A. Jalabert, and P. Brune, Europhys. Lett. \textbf{60}, 120 (2002).

\bibitem{pc24} A. M\"{u}ller-Groeling, H. A. Weidenm\"{u}ller, and C. H. Lewenkopf, Europhys. Lett. \textbf{22}, 193 (1993).

\bibitem{pc25} J. Zhong and G. M. Stocks, Nano. Lett. \textbf{6}, 128 (2006).

\bibitem{pc26} J. Zhong and G. M. Stocks, Phys. Rev. B \textbf{75}, 033410 (2007).

\bibitem{pc27} C. Y. Yang, J. W. Ding, and N. Xu, Physica B \textbf{394}, 69 (2007).

\bibitem{anlc1} H. H. Sheinfux, Y. Lumer, G. Ankonina, A. Z. Genack, G. Bartal, and M. Segev, Science \textbf{356}, 953 (2017).

\bibitem{anlc2} R. J. Elliott, J. A. Krumhansl, and P. L. Leath, Rev. Mod. Phys. \textbf{46}, 465 (1974).

\bibitem{anlc3} V. Vettchinkina, A. Kartsev, D. Karlsson, and C. Verdozzi, Phys. Rev. B \textbf{87}, 115117 (2013).

\bibitem{anlc4} D. Semmler, K. Byczuk, and W. Hofstetter,
Phys. Rev. B \textbf{81}, 115111 (2010).

\bibitem{skmprl} S. Sil, S. K. Maiti, and A. Chakrabarti, Phys. Rev. Lett. \textbf{101}, 076803 (2008).

\bibitem{pc28} G. Montambaux, H. Bouchiat, D. Sigeti, and R. Friesner, Phys. Rev. B \textbf{42}, 7647(R) (1990). 

\bibitem{new1} D. A. Browne, J. P. Carini, K. A. Muttalib, and S. R. Nagel, Phys. Rev. B \textbf{30}, 6798(R) (1984).

\bibitem{new2} J. P. Carini, K. A. Muttalib, and S. R. Nagel, Phys. Rev. Lett \textbf{53}, 102 (1984).

\bibitem{pc29} N. Yu and M. Fowler, Phys. Rev. B \textbf{45}, 11795 (1992).

\bibitem{pc30} S. K. Maiti and A. Chakrabarti, Phys. Rev. B \textbf{82}, 184201 (2010).

\bibitem{pc31} J. J. Palacios, J. Fernández-Rossier, and L. Brey, Phys. Rev. B \textbf{77}, 195428 (2008).

\bibitem{pc32} H. Kato and D. Yoshioka, Phys. Rev. B \textbf{50}, 4943 (1994).

\bibitem{pc33} A. Kambili, C. J. Lambert, and J. H. Jefferson, Phys. Rev. B \textbf{60}, 7684 (1999).

\bibitem{pc34} S. Gupta, S. Sil, and B. Bhattacharyya, Phys. Rev. B \textbf{63}, 125113 (2001).

\bibitem{deo} S. Viefers, P. Koskinen, P. Singha Deo, and M. Manninen, Physica E \textbf{21}, 1 (2004).

\bibitem{kus} J. F. Weisz, R. Kishore, and F. V. Kusmartsev, Phys. Rev. B \textbf{49}, 8126 (1994).

\bibitem{kulik} M. Iskin and I. O. Kulik, Phys. Rev. B \textbf{70}, 195411 (2004).

\bibitem{sai} Y. Saiga, D. S. Hirashima, and J. Usukura, Phys. Rev. B \textbf{75}, 045343 (2007).

\bibitem{gup1} S. Gupta, S. Sil, and B. Bhattacharyya, Physica B \textbf{355}, 299 (2005).

\bibitem{fb1} Z. Cui, J. Rothman, M. Klaui, L. Lopez-Diaz, and C. A. F. Vaz, and J. A. C. Bland, Microelectron. Eng.
\textbf{61–62}, 577 (2002).

\bibitem{fb3} M. Winzer, M. Kleiber, N. Dix, and R. Wiesendanger, Appl. Phys. A \textbf{63}, 617 (1996).

\bibitem{fb4} S. M. Weekes, F. Y. Ogrin, and W. A. Murray, Langmuir \textbf{20}, 11208 (2004).

\bibitem{pham} V. D. Pham, K. Kanisawa, and S. F\"{o}lsch, Phys. Rev. Lett. \textbf{123}, 066801 (2019).

\bibitem{keys} U. F. Keyser, C. F\"{u}hner, S. Borck, R. J. Haug, M. Bichler, G. Abstreiter, and W. Wegscheider, 
Phys. Rev. Lett. \textbf{90}, 196601 (2003).

\end{thebibliography}
\end{document}